\documentclass[usenatbib,usegraphicx,floats]{mn2e}
\usepackage{psfig}
\bibpunct{(}{)}{;}{a}{}{,}

\begin{document}
\title[The physics of pulses in GRBs: emission processes, temporal profiles and
time lags]{The physics of pulses in gamma-ray bursts: emission processes, temporal profiles and time lags.}
\author[Daigne \& Mochkovitch]{Fr\'{e}d\'{e}ric Daigne and Robert Mochkovitch\\
Institut d'Astrophysique de Paris, 98 bis bd. Arago, 75014 Paris, France}
\maketitle
\begin{abstract}
We present a simple, semi-analytical model to explain 
GRB temporal and spectral properties in the context of the internal
shock model. Each individual
pulse in the temporal profiles is produced by the deceleration 
of fast moving material by a
comparatively slower layer within a relativistic wind. 
The spectral evolution of synthetic pulses is first obtained 
with standard equipartition assumptions to estimate the post-shock 
magnetic field and electron Lorentz factor. We get 
$E_{\rm p}\propto t^{-\delta}$ with $\delta=7/2$ which is much steeper
than the observed slopes $\delta_{\rm obs}\la 1.5$. We therefore consider
the possibility that the equipartition parameters depend on the shock
strength and post-shock density. We then get a much better agreement 
with the observations and our synthetic pulses satisfy both the 
hardness-intensity and hardness-fluence correlations. We also 
compute time lags between profiles in different energy channels and 
we find that they decrease with increasing hardness. We 
finally compare
our predicted time lag -- luminosity relation to the \citet{norris:00}
result obtained from 6 bursts with  known redshift.  
\end{abstract}

\section{Introduction}
Cosmic gamma-ray burst (hereafter GRBs) exhibit a great diversity of duration
and profiles. The distribution of durations is clearly bimodal with
two peaks at about 0.2 and 20 seconds. GRB light curves are highly variable
but can often be interpreted in terms of a succession of elementary pulses which
possibly overlap \citep{norris:96}. These pulses appear as the building blocks of the 
profiles and understanding their physical origin 
would certainly re\-pre\-sent a clue for a better 
description of the whole GRB phenomenon. The pulse temporal evolution has
often been described by a fast rise followed by an exponential decay
(the so-called FRED shape; see \citet{fishman:94}) but other mathematical behaviors such as
stretched exponentials, gaussian \citep{norris:96} or power-law decays \citep{ryde:00}
have been also proposed.
Spectral hardness decreases during pulse decay and two relations between
the temporal and spectral properties, the HIC 
(hardness-intensity correlation; \citet{golenetskii:83}) and the HFC (hardness-fluence
correlation; \citet{liang:96}) appear to be satisfied by a substantial fraction
of GRB pulses during the decay phase. Pulse profiles peak earlier 
in higher energy bands and the corresponding time lags between different
energy channels correlate to pulse hardness and peak luminosity \citep{norris:00}.
These observational results must be reproduced by the models and may
help to discriminate among different possibilities.\\

Two distinct mechanisms have been proposed to explain the origin of 
pulses in GRBs. In the external shock model they are formed when a
relativistic shell ejected by the central engine is decelerated by the 
circumstellar material \citep{meszaros:93}. An homogeneous medium leads to
a single pulse but an irregular, clumpy environment can 
produce a complex profile if a large number of small 
clouds are present \citep{dermer:99}. In the internal shock model
\citep{rees:94} the central engine generates a relativistic flow with a highly
non uniform distribution of the Lorentz factor and the pulses are made 
by collisions between rapid and slower parts of the flow. In the two
scenarios the variability of the profiles has a very different 
interpretation. In one case it provides a ``tomography'' of the 
burst environment while in the second it reveals the activity of the 
central engine. \\

In this paper we consider in some details the me\-cha\-nism of pulse formation
by
internal shocks.
Three characteristic time scales may be 
relevant during pulse evolution: the time $t_{\rm rad}$ required
to radiate the energy dissipated in shocks; the dynamical time $t_{\rm dyn}$,
i.e. the time taken by internal shocks to travel throughout the flow and 
the angular spreading time $t_{\rm ang}$ cor\-res\-pon\-ding to the delay in
arrival time
of photons emitted from a spherical shell. 
A short radiative time 
$t_{\rm rad}\ll t_{\rm dyn},t_{\rm ang}$ 
appears to be mandatory to avoid adiabatic losses and maintain a sufficient
efficiency. This condition is 
satisfied by the  synchrotron process which is the most commonly invoked
radiation mechanism in GRBs.
If 
the thickness of colliding shells is small compared
to their initial separation, $t_{\rm dyn}\ll t_{\rm ang}$ and the
pulse temporal evolution is fixed by geometry; conversely if the
source produces a continuous wind rather than a series of discrete,
well se\-pa\-ra\-ted shells, $t_{\rm dyn}\ga t_{\rm ang}$
and hydrodynamical effects control the pulse shape. 
\\

Pulse evolution has been studied extensively when it is dominated 
by geometry (see e.g. \citet{fenimore:96,kobayashi:97}) but discrepancies between model predictions and
the observations \citep{soderberg:01} have cast some doubt about the validity of the internal
shock model. Our purpose is to see if the situation can be improved 
when the hydrodynamical point of view is adopted. We first summarize 
in Sect. 2 some basic informations regarding pulse temporal
and spectral evolution. We then develop in Sect. 3 a simple model
where pulses are formed when a fast moving wind is decelerated 
by a comparatively slower shell. Spectral evolution is considered
in Sect. 4 where constraints are obtained on the GRB radiation
mechanism.  
Temporal profiles computed with our model are presented in Sect. 5 and 
time lags are discussed in Sect. 6. Sect. 7 is the conclusion.  
\section{Temporal and spectral evolution during pulse decay}
We consider a pulse characterized by a photon flux $N(t)$ in the 
energy range ($E_1$, 
$E_2$), a peak energy $E_{\rm p}(t)$
of the $E^2{\cal N}(E,t)$ spectrum 
and a photon fluence defined
by
\begin{equation}
\frac{\mathrm{d}\Phi_\mathrm{N}(t)}{\mathrm{d}t} = N(t)
= \int_{E_1}^{E_2} {\cal N}(E,t)\mathrm{d}E \ .
\end{equation}

The HIC and the HFC are then given by
\begin{equation}
E_\mathrm{p}(t)\propto N(t)^{\delta}
\label{eq:HIC}
\end{equation}
and 
\begin{equation}
E_\mathrm{p}(t)\propto e^{-a \Phi_{\rm N}(t)}\ ,
\label{eq:HFC}
\end{equation}
where $a$ is an exponential decay constant.
For pulses satisfying both the HIC and the HFC, \citet{ryde:00} have shown 
that the photon flux and the peak energy follow simple power laws
during the decay phase
\begin{equation}
N(t) = \frac{N_{0}}{1+t/\tau}
\label{eq:N}
\end{equation}
and 
\begin{equation}
E_\mathrm{p}(t) = \frac{E_{\mathrm{p},0}}{(1+t/\tau)^{\delta}}\ ,
\label{eq:Ep}
\end{equation}
where $t=0$ corresponds to the maximum of $N(t)$.\\

\citet{ryde:02} performed a detailed analysis of the decay
behavior of a sample of 25 long and bright pulses to check whether 
it was indeed 
described by eqs~\ref{eq:N} and \ref{eq:Ep}. They found that 
to account for 
the temporal and spectral evolution of all the pulses,
eq.~\ref{eq:N} had to be replaced by
the more general expression
\begin{equation}
N(t) = \frac{N_{0}}{(1+t/\tau)^{n}}\ .
\label{eq:Nbis}
\end{equation}
  
If $n\ne 1$, the $\delta$ indices appearing in eqs~\ref{eq:HIC} and
\ref{eq:Nbis} are different
and following \citet{ryde:02} we then write 
$E_\mathrm{p}(t) ={E_{\mathrm{p},0}}/{(1+t/\tau)^{\delta_{\star}}}$
with $\delta_{\star}=n\delta$.
\citet{ryde:02} found that
the distribution of $n$ in their sample 
was sharply peaked at $n=1$ with however a secondary bump at $n\la 3$.
The values of 
$\delta_{\star}$
were all smaller than 1.5 for the $n=1$ pulses but could reach 3.5 when
$n\approx 3$ . The distribution of $\delta$ was narrower with
$0.5\la \delta \la 1$ in most of the sample.\\
\begin{figure}
\centerline{\psfig{figure=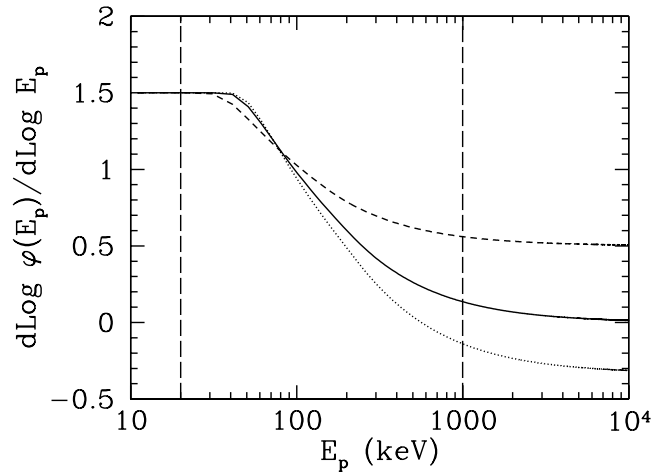,width=0.49\textwidth,angle=-90}}
\caption{Derivative of $\varphi(E_{\rm p})$ for three 
Band functions with $\beta=-2.5$ and $\alpha=-2/3$ (dotted line)
$\alpha=-1$ (full line) and $\alpha=-1.5$ (dashed line). 
The two vertical lines limit the BATSE spectral range.
The average 
slope during pulse decay typically lies between 0 and 1.}
\label{fig:slope}
\end{figure}
   
Once the decay behavior of 
$N(t)$ and $E_\mathrm{p}(t)$ has been specified, it becomes possible to obtain
the
evolution of 
the bolometric energy flux $F_\mathrm{E}(t)$
since
\begin{equation}
F_\mathrm{E}(t) = \int_0^{\infty}\mathcal{N}(E,t)E \mathrm{d}E 
= E_\mathrm{p}^2 \int_0^{\infty}\mathcal{N}(x,t)x \mathrm{d}x\ ,
\end{equation}
where $x=E/E_\mathrm{p}$.
We suppose that the temporal and spectral behavior can be separated in
$\mathcal{N}(x,t)$ :
\begin{equation}
\mathcal{N}(x,t) = A(t)\mathcal{B}(x)\ ,
\end{equation}
$\mathcal{B}(x)$ representing the spectrum shape. The photon flux in the energy
range
$(E_1,E_2)$ is then given by
\begin{equation}
N(t) =
\int_{E_1}^{E_2}\mathcal{N}(E,t)\mathrm{d}E=\frac{F_\mathrm{E}(t)
\varphi(E_\mathrm{p})}{E_\mathrm{p}(t) \varphi_0}\ ,
\label{eq:Ndef}
\end{equation}
so that 
\begin{equation}
F_\mathrm{E}(t)=N(t)E_{\rm p}(t)\;\frac{\varphi_0}{\varphi(E_\mathrm{p})}
={N_0 E_{{\rm p},0}\over (1+t/\tau)^{1+\delta_{\star}}} \;
\frac{\varphi_0}{\varphi(E_\mathrm{p})}
\label{eq:FE}
\end{equation}
with
\begin{equation}
\varphi(E_\mathrm{p})=\int_{E_1/E_\mathrm{p}}^{E_2/E_\mathrm{p}}
\mathcal{B}(x)\mathrm{d}x
\end{equation}
and
\begin{equation}
\varphi_0 = \int_0^{\infty} x \mathcal{B}(x)\mathrm{d}x\ .
\end{equation}

The derivative of $\varphi(E_{\rm p})$ has
been represented in Fig.~\ref{fig:slope} for the BATSE spectral range (20, 1000
keV),
using a standard Band function \citep{band:93} with low and high
energy indices $\alpha=-2/3$, $-1$ or $-1.5$ and $\beta=-2.5$. 
At low (resp. high) $E_{\rm p}$, $\varphi(E_{\rm p})$ is given by a
simple power-law $E_{\rm p}^{-(\beta+1)}$ (resp. $E_{\rm p}^{-(\alpha+1)}$)
but for intermediate values 
($E_1<E_{\rm p}<E_2$) which are representative of the decay phase in the
\citet{ryde:02} sample, $\varphi(E_{\rm p})$ does not have a simple
ana\-ly\-ti\-cal form. Assuming that it can still be approximated 
by a power-law,
$\varphi(E_{\rm p})\propto E_{\rm p}^{-(\zeta+1)}$ where $\zeta$ 
is a weighted average of the low and high energy spectral indices
($-2\la \zeta\la -1$) the bolometric energy flux also
follows a power-law   
\begin{equation}
F_\mathrm{E}(t) \propto \frac{1}{(1+t/\tau)^{\epsilon}}
\label{eq:NewFE}
\end{equation}
with 
\begin{equation}
\epsilon=n+(2+\zeta)\,\delta_{\star}\ .
\label{eq:epsilon}
\end{equation} 
The slope of the HIC is then given by
\begin{equation}
\delta={\delta_{\star}\over n}={1\over \epsilon/\delta_{\star}
-(2+\zeta)}\ .
\end{equation}

If the temporal and spectral evolution during pulse decay is due to
geometrical effects alone $\epsilon=3$ and $\delta_{\star}=1$ \citep{granot:99} which leads
to
$\delta={1\over 1-\zeta}$. With 
$-2\la \zeta\la -1$
the resulting value $0.3\la\delta\la 0.5$ lies below what is found in most 
observed pulses \citep{soderberg:01}.\\

Geometrical effects govern pulse evolution if the shell
thickness is small compared to their initial separation. 
But if a continuous  
outflow emerges from the central engine the hydrodynamical time scale
can play a dominant role du\-ring pulse decay. We have then developed 
a simple model to check whether a better agreement can 
be found
with the observations when the hydro\-dy\-na\-mi\-cal
aspect of the flow is taken into account.  
\begin{figure}
\centerline{\psfig{figure=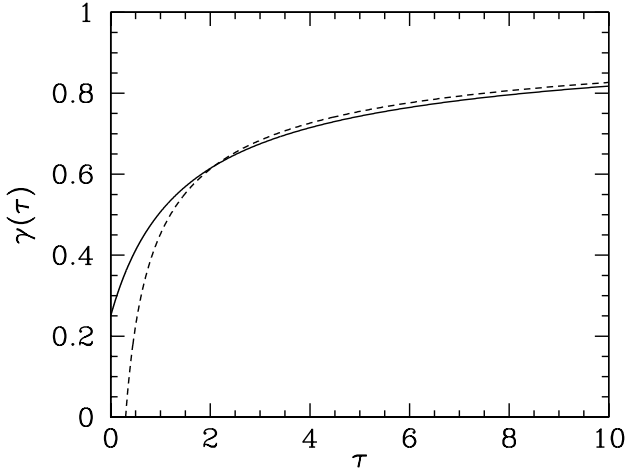,width=0.49\textwidth,angle=-90}}
\caption{Solution for $\gamma(\tau)$ corresponding to eq.~\ref{eq:solution} with
$\gamma_0=0.25$. The dashed line is the approximation given by
eq.~\ref{eq:approx}.}
\label{fig:gamma}
\end{figure}
\begin{figure}
\centerline{\psfig{figure=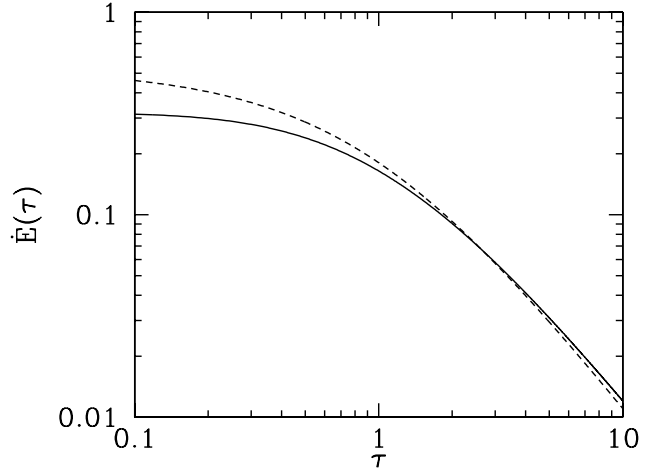,width=0.49\textwidth,angle=-90}}
\caption{Dissipated power in units of ${{\dot M}\Gamma_1 c^2\over 2}$
for our pulse model with
$\gamma_0=0.25$. The dashed line corresponds to eq.~\ref{eq:power} 
while the full line
takes into account angular spreading.}
\label{fig:power}
\end{figure}

\section{A simple pulse model} 
We consider a relativistic wind where a slow shell of mass $M_0$ and Lorentz
factor $\Gamma_0$ decelerates a more rapid part of the flow characterized by
a constant mass flux $\dot{M}$ (in the source frame) and Lorentz factor 
$\Gamma_1>\Gamma_0$. 
We do not solve the true hydrodynamical problem
but rather approximate the flow evolution by considering that fast material 
is ``accreted'' by the slow shell. The accretion rate is given by
\begin{equation}
\frac{\mathrm{d}M}{\mathrm{d}t} = \dot{M}(1-\gamma^2)\ ,
\label{eq:mdot}
\end{equation}
where $t$ is the observer time and $\gamma=\Gamma/\Gamma_1$ 
($\Gamma$ and $M$ being the current Lorentz factor and mass of the slow
shell). Due to the accretion of fast moving material, the Lorentz factor 
of the slow shell increases. When a mass element $\mathrm{d}M$ is accreted the
Lorentz factor
becomes
\begin{equation}
\Gamma+\mathrm{d}\Gamma=\left(\Gamma_1\Gamma {\Gamma_1 \mathrm{d}M+\Gamma M\over
\Gamma\mathrm{d}M+\Gamma_1 M}\right)^{1/2}\ ,
\end{equation}
so that
\begin{equation}
\frac{\mathrm{d}\gamma}{\mathrm{d}M} = \frac{1-\gamma^2}{2 M}\ ,
\end{equation}
which can be integrated to give 
\begin{equation}
\mu=\left(\frac{1+\gamma}{1-\gamma}\right)/\left(\frac{1+\gamma_0}
{1-\gamma_0}\right)\ ,
\label{eq:mu}
\end{equation}
where $\mu=M/M_0$ and $\gamma_0=\Gamma_0/\Gamma_1$. Introducing 
$t_0=M_0/\dot{M}$ and $\tau=t/t_0$, eqs~\ref{eq:mdot}--\ref{eq:mu}  yield
\begin{equation}
\frac{\mathrm{d}\gamma}{\mathrm{d}\tau} = Q(1-\gamma^2)(1-\gamma)^2
\label{eq:equadiff}
\end{equation}
with $Q=\frac{1}{2}\left(\frac{1+\gamma_0}{1-\gamma_0}\right)$.
Equation \ref{eq:equadiff} has the analytical solution
\begin{equation}
\tau=\frac{1}{Q}\left[F(\gamma)-F(\gamma_0)\right]\ ,
\label{eq:solution}
\end{equation}
where the function $F(\gamma)$ is given by
\begin{equation}
F(\gamma)=\frac{1}{8}\log{\left(\frac{1+\gamma}{1- \gamma}\right)}
+\frac{1}{4(1-\gamma)}+\frac{1}{4(1-\gamma)^2}\ .
\end{equation}

The solution $\gamma(\tau)$ corresponding to eq.~\ref{eq:solution} has been
represented in
Fig.~\ref{fig:gamma} for $\gamma_0=0.25$. When $\tau\ga 2$, it is well 
ap\-pro\-xi\-ma\-ted by
\begin{equation}
\gamma(\tau) \simeq 1-\frac{1}{2\sqrt{Q\tau}}\ .
\label{eq:approx}
\end{equation}

Once $\gamma(\tau)$ is known it is possible to calculate the dissipated power
\begin{equation}
\dot{\mathcal{E}}(\tau) = \frac{\dot{M}\Gamma_1 c^2}{2}
(1-\gamma^2)(1-\gamma)^2\ ,
\label{eq:power}
\end{equation}
which has been represented in Fig.~\ref{fig:power}. At large $\tau$, 
it behaves as $\tau^{-3/2}$ since 
\begin{equation}
\dot{\mathcal{E}}(\tau)\propto (1-\gamma)^3 (1+\gamma)\propto \tau^{-3/2}
\label{eq:approxpower}
\end{equation}
for $\tau\ga 2$. 
Pulse evolution is essentially completed at $\tau\sim 10$ when
$\Gamma/\Gamma_1>0.8$ and $\dot{\mathcal{E}}$ has decreased by more than
an order of magnitude.\\

The dissipated power given by eq.~\ref{eq:power} is slightly 
different from what the observer will see
since the energy released at time $t$ is 
spread over an interval $\Delta t$ corresponding to the difference in arrival 
time for photons emitted by a shell of radius $r$ moving at a Lorentz
factor $\Gamma$
\begin{equation}
\Delta t=\frac{r}{2 c \Gamma^2}\ .
\end{equation}

The solution for $\dot{\mathcal{E}}$ including angular spreading has been
obtained numerically and is also shown in Fig.~\ref{fig:power}. It differs from
the
analytical expression (eq.~\ref{eq:power}) at early times but preserves the 
power law decay of slope $\epsilon=3/2$ at late times. 
\begin{figure}
\psfig{figure=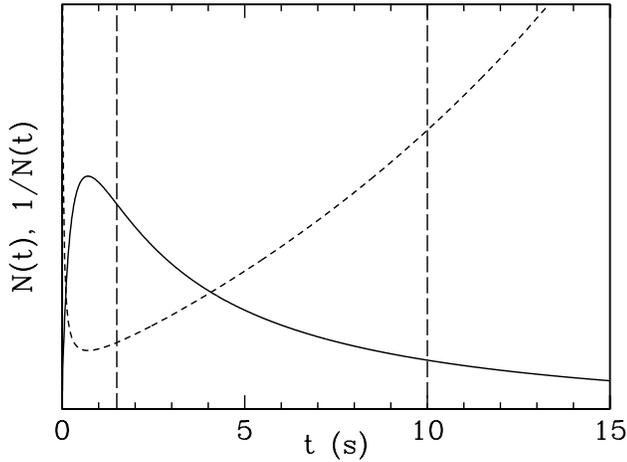,width=0.49\textwidth,angle=-90}
\caption{Pulse profile $N(t)$ (full line) and $1/N(t)$ (dashed line)
produced by a relativistic outflow with
$\Gamma_1=400$ decelerated by a slow shell of initial Lorentz factor
$\Gamma_0=100$ (see text for the other model parameters). The dashed 
vertical lines limit the time interval where we plot the HIC and 
the HFC in Fig.~\ref{fig:timelag}.}
\label{fig:profiles}
\end{figure}
\begin{figure}
\psfig{figure=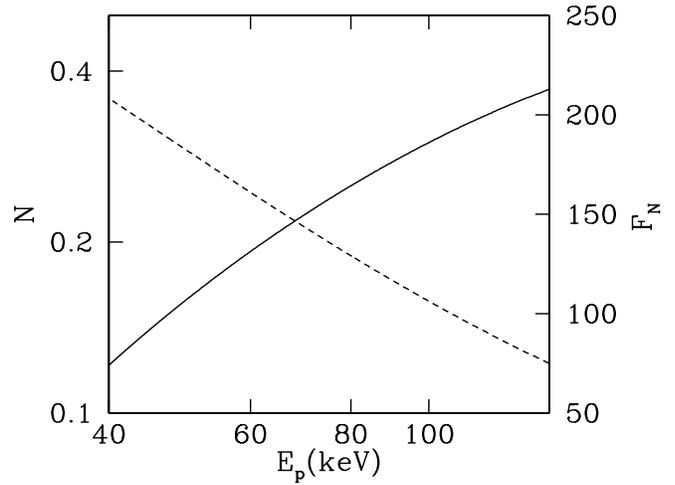,width=0.49\textwidth,angle=-90}
\caption{Hardness-intensity (full line) and hardness-fluence 
(dashed line) correlations for the pulse
shown in Fig.~\ref{fig:profiles}. }
\label{fig:HICHFC}
\end{figure}

\section{Spectral evolution and emission processes} 
We now use the analytical model to follow the spectral evolution
during pulse decay.
If the dissipated energy is radiated by the synchrotron process the 
peak energy $E_\mathrm{p}$ is
\begin{equation}
E_\mathrm{p} = E_\mathrm{syn} \propto \Gamma \,B \Gamma_\mathrm{e}^2 \ ,
\end{equation}
where $B$ is the magnetic field and $\Gamma_\mathrm{e}$ the characteristic
electron Lorentz
factor behind the shock. 
With classical equipartition assumptions $B$ and $\Gamma_\mathrm{e}$ can be
expressed
as 
\begin{equation}
B = (8\pi\alpha_\mathrm{B} \rho \epsilon\,c^2)^{1/2}
\end{equation}
and
\begin{equation}
\Gamma_\mathrm{e}=\frac{\alpha_\mathrm{e}}{\zeta}\frac{m_\mathrm{p}}{m_\mathrm{e
}}\epsilon\ ,
\end{equation}
where $\rho$ is the density and $\epsilon c^2$ the 
dissipated energy per unit
mass (both in the comoving frame); $\alpha_\mathrm{B}$ and $\alpha_\mathrm{e}$ are the
equipartition 
parameters and $\zeta$ is the fraction of electrons which are accelerated.
Finally,
\begin{equation}
E_\mathrm{syn} \propto \Gamma\,\rho^{1/2} \epsilon^{5/2}\ ,
\label{eq:Esyn}
\end{equation}
where the comoving density $\rho$ is proportional to $r^{-2}$ 
($r$ being the shock radius
$r\sim \Gamma^2 c t$) and $\epsilon$ is obtained from 
$\dot{\mathcal{E}}={dM\over dt}\Gamma\epsilon c^2$ and eq.~\ref{eq:mdot}
\begin{equation}
\epsilon=\frac{(1-\gamma)^2}{2 \gamma}\ .
\end{equation}

This leads to the following expression for $E_\mathrm{syn}$ 
\begin{equation}
E_\mathrm{syn} \propto \frac{(1-\gamma)^5}{\gamma^{7/2}t}\ ,
\end{equation}
which behaves as a power law ($E_\mathrm{p}\propto t^{-7/2}$) 
when $(1-\gamma)\sim t^{-1/2}$.
This is much steeper than the observed spectral evolution 
of pulses which satisfy both the HIC and the HFC.
Instead of using eq.~\ref{eq:Esyn} we therefore parametrize the peak energy   
with the more general phenomenological expression
\begin{equation}
E_\mathrm{p} \propto \Gamma\,\rho^x \epsilon^y  \propto
\frac{(1-\gamma)^{2y}}{\gamma^{4x+y-1}t^{2x}}\ ,
\label{eq:Epxy}
\end{equation}
which becomes
\begin{equation}
E_\mathrm{p} \propto \frac{1}{t^{2x+y}}
\label{eq:EpApprox}
\end{equation}
at late times.
The exponents $x$ and $y$ can be different from their standard synchrotron 
values $1/2$ and $5/2$ if the equipartition parameters 
$\alpha_\mathrm{B}$, $\alpha_\mathrm{e}$ or $\zeta$ vary with $\rho$ or/and
$\epsilon$. 
For example, \citet{daigne:98} adopted a fraction $\zeta$ of
accelerated electrons proportional to $\epsilon$ so that $\Gamma_\mathrm{e}$ 
remains constant which leads to $x=y=1/2$ and $E_\mathrm{p}\propto t^{-3/2}$.
However, since most of the observed values of $\delta_{\star}=2x+y$
are smaller than 1.5, it seems necessary to further reduce the
$x$ and $y$ indices and we have therefore considered below 
the case $x=y=1/4$, i.e. $\delta_{\star}= 0.75$.

\section{Temporal profiles} 
We obtain the temporal profile of synthetic pulses  
from eqs~\ref{eq:Ndef}, \ref{eq:power} and \ref{eq:Epxy} of our model. 
We have represented in Fig.~\ref{fig:profiles} a pulse 
formed when a wind of Lorentz factor $\Gamma_1=400$ and 
power $\dot{M}\Gamma_1 c^2=10^{52}$ erg.s$^{-1}$ 
is decelerated by a slow shell with 
$\Gamma_0=100$. We adopt $t_0=0.4$ s, $x=y=1/4$ and $z=1$.  
The profile 
is computed in the BATSE range
(20 -- 1000 keV) and the constant of proportionality 
in eq.~\ref{eq:Epxy} is fixed
to get a peak energy $E_\mathrm{p}=300$ keV
for the whole pulse spectrum. 
The pulse duration is close to $10\,(1+z)\,t_0$ as expected from the
results obtained in Sect.3.

The evolution after maximum is initially close
to a $1/t$ decay (i.e. $n\sim 1$ in eq.~\ref{eq:Nbis}) as can be 
seen in Fig.~\ref{fig:profiles} where $1/N(t)$ has been also represented.
This can be simply understood from eq.~\ref{eq:epsilon} which, 
for the decay slopes of the dissipated power $\epsilon=1.5$
(eq.~\ref{eq:approxpower})
and of the peak energy 
$\delta_{\star}= 2x+y=0.75$ gives
\begin{equation}
n=-0.75\,\zeta\ .
\end{equation}
 
With $-2 \la \zeta\la -1$, the central value of
$n$ is indeed close to unity.
Since the decay phase of our synthetic pulse can be  
decribed by eqs.~\ref{eq:N} and \ref{eq:Ep} it should also satisfy 
both the HIC and the
HFC. This is checked in Fig.~\ref{fig:HICHFC} where the two relations have been
plotted
in the time interval delimited by the two vertical lines in
Fig.~\ref{fig:profiles}.
The HIC is not a strict power law but its average slope $\delta\sim 0.9$
is 
in global agreement with the observations. The HFC is satisfied to a
better
accuracy since the relation between $\log\,E_{\rm p}$ and the photon 
fluence is quasi-linear in the considered interval.  
\begin{figure}
\resizebox{0.45\textwidth}{!}{\mbox{\includegraphics*[1.75cm,15cm][17.75cm,28cm]
{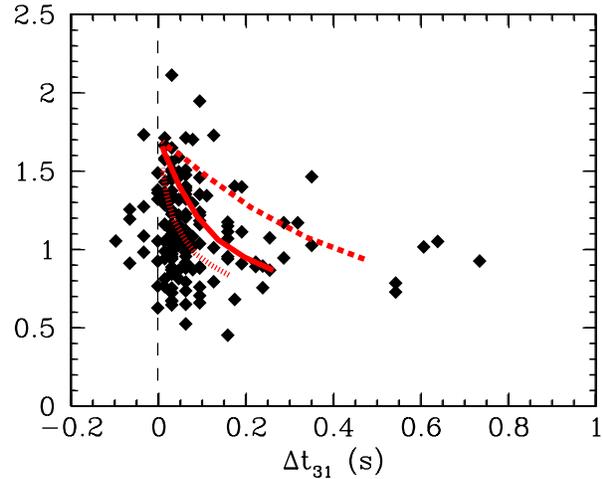}}}
\caption{Time lag -- hardness ratio diagram of BATSE bursts from 
\citet{norris:00} 
compared to model predictions. The three lines 
represent sequences of synthetic pulses obtained
with similar distributions of the Lorentz factor 
(the dotted line corresponds to $\Gamma_1=300$, the full line to
$\Gamma_1=400$ and the dashed line to $\Gamma_1=600$; see text for
details).  }
\label{fig:timelag}
\end{figure}

\section{Time lags} 
\citet{norris:00} have shown that time lags between different energy 
channels correlate with spectral hardness 
and possibly also with the burst peak luminosity. GRBs 
are distributed in a
triangular domain of the time lag--hardness ratio diagram 
(the hardest bursts having the shortest time lags; see Fig.~\ref{fig:timelag})
and 
the time lag--luminosity relation obtained from 6 bursts with known
redshifts takes the form
\begin{equation}
L_{51} \simeq 130\left(\frac{\Delta t_{31}}{0.01\ \mathrm{s}}\right)^{-1.14}\ ,
\label{eq:L51}
\end{equation}
where $\Delta t_{31}$ is the time lag between BATSE channels 3 and 1 and 
$L_{51}$ is the luminosity in units of $10^{51}$ erg.s$^{-1}$.  
In our model, we estimate time lags by cross-correlating profiles in different
energy
channels obtained from eq.~\ref{eq:Ndef}. 
The first factor in eq.~\ref{eq:Ndef} is 
\begin{equation}
\psi(t)=\frac{F_\mathrm{E}(t)}{E_\mathrm{p}(t)}\ ,
\end{equation}
which behaves as $t^{2x+y-1.5}$ during pulse decay. 
The sign of $\Delta=2x+y-1.5$ is of great importance in determining the time
$t_\mathrm{max}$ of maximum count rate and the related time lags. If $\Delta<0$
the function $\psi(t)$ has a maximum at some early
time $t_\mathrm{m}$ before decreasing as $t^{\Delta}$ while it steadily
increases
for $\Delta>0$.
The second factor in eq.~\ref{eq:Ndef} is $\varphi(E_\mathrm{p})$ so that
$t_\mathrm{max}$ is solution of the implicit equation  
\begin{equation}
{{\dot \psi}(t_{\rm max})\over \psi(t_{\rm max})}+
{{\dot E}_{\rm p}^{\rm max}\over E_{\rm p}^{\rm max}}
{\mathrm{d}\log\,\varphi\over \mathrm{d}\log\,E_{\rm p}}\Big|_{E_{\rm p}^{\rm
max}}=0\ ,
\label{eq:tmax}
\end{equation} 
where $E_\mathrm{p}^\mathrm{max}$ 
and 
${\dot E}_\mathrm{p}^\mathrm{max}$ 
are the values of $E_\mathrm{p}$ and its time derivative
at $t=t_\mathrm{max}$.
Since in most pulses the evolution of $E_{\rm p}$ precedes the count rate, 
${\dot E}_{\rm p}^{\rm max}$ is negative. If the low energy 
slope of the Band spectrum $\alpha\le -1$, the derivative   
${\mathrm{d}\log\,\varphi\over \mathrm{d}\log\,E_{\rm p}}\Big|_{E_{\rm p}^{\rm
max}}$ 
is positive (see Fig.~\ref{fig:slope}) and  
eq.~\ref{eq:tmax} then shows that $\dot{\psi}(t_\mathrm{max})>0$  
which, for $\Delta<0$, leads to  
\begin{equation}
t_\mathrm{max}<t_\mathrm{m}\ ,
\label{eq:tmaxm}
\end{equation}
which provides a strict upper limit on the time lags
between 
different energy channels
\begin{equation}
\Delta t<t_{\rm m}\ .
\end{equation} 
(If $\alpha>-1$,  
${\mathrm{d}\log\,\varphi\over \mathrm{d}\log\,E_{\rm p}}\Big|_{E_{\rm p}^{\rm
max}}$
can be weakly negative at large $E_{\rm p}$ 
but even in this case $t_{\rm max}$ never greatly 
exceeds $t_{\rm m}$.)\\
                               
When $\Delta \ge 0$, no constraint such (40)
applies and the time lags can be quite large.
\citet{daigne:98} who adopted $x=y=1/2$ ($\Delta=0$) obtained     
lags of several seconds between BATSE channels 1 and 3 (or 4) for a pulse
lasting about 10 s, while currently observed values are in the range
$10^{-2}$ to a few $10^{-1}$ s. 
Moreover,                                          
for $\Delta>0$ the time lags increase with pulse hardness, in contradiction
with the observations.\\

Conversely, with $x=y=1/4$ the time lags are short 
($\Delta t_{31}\la 0.5$ s) even for long pulses and 
they decrease with increasing hardness. Figure~\ref{fig:timelag} shows 
the time lag -- hardness ratio relation given by our model superimposed 
to the 
\citet{norris:00} results for BATSE bursts.
The thick grey lines in Fig.~\ref{fig:timelag}  
correspond to sequences of pulses of comparable duration 
($t_{90}\sim 10$ s) 
obtained with similar distributions 
of the Lorentz factor (a slow shell with $\Gamma_0=100$ decelerating a fast 
wind with $\Gamma_1=300$, 400 or 600)
and a varying value
of the 
(isotropic) injected
power (from 
$5\ 10^{51}$ to $10^{54}$ erg.s$^{-1}$).
Even if differences in duration and redshift will contribute to 
add scatter to their distribution it appears that synthetic pulses
populate the same triangular domain as observed ones.\\

\begin{figure}
\resizebox{0.45\textwidth}{!}{\mbox{\includegraphics*[0.75cm,15cm][17.75cm,28cm]
{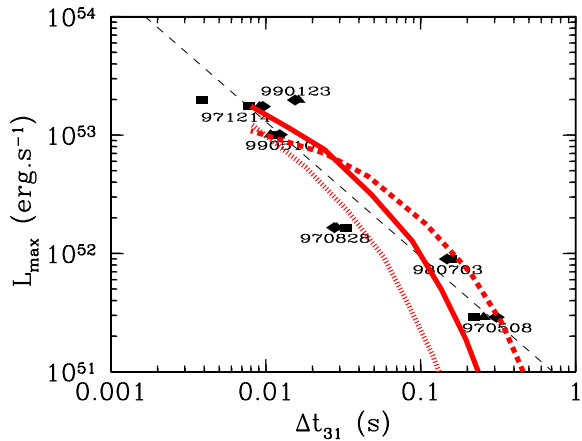}}}
\caption{Time lag -- luminosity correlation predicted by our model compared
to \citet{norris:00} results for 6 GRBs with known redshifts.
The lines correspond to the three cases already considered in
Fig.~\ref{fig:timelag}.}
\label{fig:L51}
\end{figure}

We finally checked if our model was able to reproduce the time lag -- 
luminosity correlation (eq.~\ref{eq:L51}). 
When $x=y=1/4$, we do find that the lags decrease with increasing 
luminosity. This is a consequence of the HIC and the time lags --
hardness ratio relation discussed above. 
The results are shown in Fig.~\ref{fig:L51} where the three lines correspond
to the wind cases with $\Gamma_1=300$, 400 and  600
already considered in Fig.~\ref{fig:timelag}. It can be seen that there is an
overall
agreement between the model predictions and eq.~\ref{eq:L51}. However, at low
luminosities and large time lags we obtain a rather wide strip instead 
of a single relation as eq.~\ref{eq:L51}. If this is confirmed by the 
analysis of more GRBs with known redshifts, the time lag -- luminosity
correlation will still be useful for statistical studies of large burst
samples but may be quite inaccurate to estimate the luminosity 
of a specific event.    

\section{Conclusion} 
We have developed a simple model where GRB pulses are produced when a rapid
part of a relativistic outflow is dece\-le\-ra\-ted by a comparatively slower 
shell. We do not solve the true hydrodynamical problem but rather assume 
that the slow shell ``accretes'' the fast moving material which allows to
obtain an analytical solution for the dissipated power $\dot{\mathcal{E}}$.
During pulse decay $\dot{\mathcal{E}}\propto t^{-3/2}$ as
$\dot{\mathcal{E}}\propto t^{-3}$ when the evolution is fixed by 
shell geometry.
To compute the spectral evolution of our synthetic pulses we 
parametrize the peak energy as $E_\mathrm{p} \propto \rho^x \epsilon^y \Gamma$
where $\rho$, $\epsilon c^2$ and $\Gamma$ are respectively the post shock 
values of the density, dissipated energy (per unit mass) and Lorentz factor. 
At late times, we get $E_\mathrm{p}\propto t^{-(2x+y)}$ which 
constraints $x$ and
$y$ since in most observed bursts $E_\mathrm{p} \propto t^{-\delta_{\star}}$ 
with $\delta_{\star}\la 1.5$. 
The synchrotron process with standard equipartition assumptions corresponds
to $x=1/2$ and $y=5/2$ (i.e. $2x+y=3.5$) and gives a much too steep
spectral evolution.
One has then to suppose that the equipartion parameters 
$\alpha_\mathrm{e}$, $\alpha_\mathrm{B}$
and $\zeta$ vary with $\rho$ or/and $\epsilon$ to reduce $x$ and $y$
(a possible alternative being that energy is
radiated by another process -- different from synchrotron --  
but which can still be approximated by eq.~\ref{eq:Epxy}). \\

We have considered
the case $x=y=1/4$ 
and the resulting pulses then have temporal and spectral properties 
in excellent agreement with the observations. They follow both the
HIC and the HFC during the decay phase and the time lags between
energy channels decrease with increasing pulse hardness and peak luminosity.
We therefore conclude that if GRB pulses are produced by internal
shocks, their temporal and spectral properties are probably governed by 
the hydrodynamics of the flow 
rather by the geometry of the emitting shells.

\bibliographystyle{mn2e}
\bibliography{pulses}

\end{document}